\documentstyle[epsfig,twocolumn]{mn}

\title[Milky Way rotation curve in HL theory]{The Milky Way rotation curve in Horava\,-\,Lifshitz theory}
\author[V.F. Cardone et al.]{V.F. Cardone$^{1,2}$, N. Radicella$^{3,4}$, M.L. Ruggiero$^{3,4}$, M. Capone$^{4,5}$\\
$^1$ Dipartimento di Scienze e Tecnologie dell' Ambiente e del Territorio, Universit\`{a} degli Studi del Molise, \\
Contrada Fonte Lappone, 86090\,-\,Pesche (IS), Italy \\
$^2$Dipartimento di Scienze Fisiche, Universit\`{a} degli Studi di Napoli "Federico II",
Complesso Universitario \\ di Monte Sant' Angelo, Edificio N, via Cinthia, 80126 - Napoli, Italy \\
$^3$Dipartimento di Fisica, Politecnico di Torino, Corso Duca degli Abruzzi 24, 10129 - Torino, Italy \\
$^4$I.N.F.N.\,-\,Sezione di Torino, via Pietro Giuria 1, 10125\,-\,Torino, Italy \\
$^5$Dipartimento di Matematica, Universit\`{a} degli Studi di Torino, Via Carlo Alberto 10,  10125\,-\,Torino, Italy \\}

\date{Accepted xxx, Received yyy, in original form zzz}

\begin{document}
\maketitle

\begin{abstract}

The Horava\,-\,Lifshitz (HL) theory has recently attracted a lot of interest as a viable solution
to some quantum gravity related problems and the presence of an effective cosmological constant
able to drive the cosmic speed up. We show here that, in the weak field limit, the HL proposal leads
to a modification of the gravitational potential because of two additive terms (scaling respectively
as $r^2$ and $r^{-4}$) to the Newtonian $1/r$ potential. We then derive a general
expression to compute the rotation curve of an extended system under the assumption that the mass density
only depends on the cylindrical coordinates $(R, z)$ showing that the HL modification induces a dependence
of the circular velocity on the mass function which is a new feature of the theory. As a first exploratory
analysis, we then try fitting the Milky Way rotation curve using its visible components only in order
to see whether the HL modified potential can be an alternative to the dark matter framework. This turns out
not to be the case so that we argue that dark matter is still needed, but the amount of dark
matter and the dark halo density profile have to be revised according to the new HL potential.

 \end{abstract}


\section{Introduction}

Inspired by the Lifshitz theory in condensed matter physics, Horava (2009a,b) has recently proposed a new theory of gravity
based on an anisotropic scaling of space and time in the UV limit. Usually referred to as Horava\,-\,Lifshitz (hereafter, HL) theory,
the HL proposal shows a reduced invariance, dubbed {\it foliation preserving diffeomorphism invariance}, which however reduces to the
standard one in the IR limit where General Relativity is recovered. As an attractive feature, the HL theory turns out to be power counting
renormalizable which has motivated the great interest in investigating with great detail its theoretical and cosmological aspects. It
is worth remembering that, in its original formulation, two conditions were imposed in order to drive the choice of the field action.
First, the {\it projectability condition} was supposed to hold true. Since time plays a fundamental role from the very
beginning, it was assumed that the lapse function (defined when working in the ADM formulation of gravity) has to be a projectable function
on the spacetime foliation, that is to say a function of time only. Second, in order to reduce the number of independent terms entering
the action, the principle of {\it detailed balance} was used. Unfortunately, it soon became clear that this second condition leads to problems
in the low energy limit \cite{LMP09} thus motivating the search for modification of the original HL theory where the action breaks the detailed
balance condition either softly \cite{KS09,LKM09,CP09,KK09a,KK09b} or not. In particular, Sotiriou, Visser and Weinfurtner (2009a,b) have worked
out a modified HL theory with no detailed balance condition imposing that only parity preserving operators enter the potential.

Although the discussion about the foundations and the possible conceptual and phenomenological problems of the HL theory and its modified
versions is still open, it is nevertheless worth systematically investigating its consequences at every scale. In particular, it is
interesting to study its static spherically symmetric solutions since one can thus derive the gravitational potential generated from a point
mass source. Recently, this problem has been addressed by Tang \& Chen (2009) for the HL theory with the projectability condition and no
detailed balance. They argue that only the Minkowski or de Sitter spacetime are solutions, but we will show here that this is actually not the
case. As a consequence, we find that the gravitational potential generated by a point mass differs from the Newtonian one because of the
presence of additional terms depending on the HL coupling parameters. Such terms have to be taken into account when computing the potential
generated by an extended system, such as a galaxy. Therefore, we work out a general formalism to estimate the rotation curve (i.e., the circular
velocity $v_c$ as function of the distance $R$ from the centre) for an extended source showing that the HL theory can boost the $v_c(R)$
with respect to the Newtonian value. Motivated by this consideration, we then try to fit the Milky Way rotation curve using visible matter only
to both constrain the HL parameters and investigating whether it can work as an effective dark matter component.

The plan of the paper is as follows. In Sect.\,2, we look for static spherically symmetric solutions for the HL theory with projecatibility condition
and show that there is indeed a new solution leading to a modified gravitational potential. A general formalism to compute the rotation
curve for an extended system is presented in Sect.\,3 and then used in Sect.\,4 to work out the predicted rotation curve for the Milky Way. Here, we also present the data and the results of fitting them with our modified potential and no dark matter. Conclusions are finally given in Sect.\,5.

\section{The point mass potential}\label{The point mass gravitational potential}

The weak field limit of the HL theory as proposed by Sotiriou, Visser and Weinfurtner has been yet discussed by Tang \& Chen (2009, hereafter TC09)
so that here we will only summarize the main steps and stress where our work differs from their one.
As stressed in TC09, we start from the observation that not all the static spherically symmetric solutions
found for (modifications of) HL theory preserve the projectability condition. In fact, if one assumes a metric

\begin{equation}
ds^2 = -N^2(r) dt^2 + \frac{dr^2}{g(r)} + r^2 d\Omega^2
\end{equation}
with $d\Omega^2 = d\theta^2 + \sin{\theta}^2 d\phi^2$ and use the coordinate transformation
$dt = dt_{PG} - [(1- N^2)^{1/2}/N^2]dr$, one can show
that a necessary condition for the projectability condition to hold is that $g = N^2$. The line
element therefore reads\,:

\begin{equation}
ds^2 =  -N^2(t) dt^2 + \frac{\left ( dr + N^r dt \right )
\left ( dr + N^r dt \right )}{f}  + r^2 d\Omega^2 \ .
\end{equation}
The equations of motion for such a metric are given in TC09 and will not be repeated here for sake of
shortness. The authors then argue that in the IR limit the $f$ function is constrained
to be 1 thus ending up with the usual Schwarzschild\,-\,de Sitter metric as only solution. We here show
that this is actually not the case. To this aim, we first note that the Euler\,-\,Lagrange equation for
$N_r = N^r/f$ in the IR limit reduces to

\begin{equation}
\frac{f'}{f}\frac{N_r}{r} = 0
\label{Nr}
\end{equation}
which is solved by either $N_r = 0$ or $f = const$. The first choice gives Minkowski back when choosing $g_0 = 0$ and  $f=1$,
or simply reduces to Minkowski by means of a redefinition of the radial coordinate. The second choice seems therefore more
interesting. In such a case, the equations obtained by varying with respect to $f$ and $N$ reduce to

\begin{equation}
\frac{dN_r^2}{dr} + \frac{N_r^2}{r} + \frac{N^2(t)}{2f^2} \nu_1(r) = 0  \ ,
\label{eq: ss1}
\end{equation}

\begin{equation}
\int_{0}^{\infty}{\left [ \frac{dN_r^2}{dr} + \frac{N_r^2}{r} + \frac{N^2(t)}{2f^2} \nu_2(r) \right ] r^3 dr} = 0 \ ,
\label{eq: ss2}
\end{equation}
with

\begin{eqnarray}
\nu_1(r) & = & -g_{0} \zeta^{6} r + \frac{2 \zeta^{4}(1 - f)}{r} \nonumber \\
~ & + & \frac{2\zeta^{2}(1 - f)}{r^3} \left [ 2g_{2}(1 + 7f) + g_{3}(1 + 5f) \right ] \nonumber \\
~ & + & \frac{2(1 - f)^2}{r^5} \left [ 4g_{4}(1 + 23f) + 2g_{5}(1 + 17f) \right . \nonumber \\
~ & + & \left . g_{6}(1 + 14f) \right ] + \frac{8f(1 - f)}{r^5} \left [ 2g_{7}(1 + 7f) \right . \nonumber \\
~ & + & \left . g_{8}(1 - 4f) \right] \ ,
\label{eq: defnu1}
\end{eqnarray}

\begin{eqnarray}
\nu_2(r) & = & -g_{0} \zeta^{6} r + \frac{2\zeta^{4}(1 - f)}{r} \nonumber \\
~ & + & \frac{2\zeta^{2}(1 - f)^2}{r^{3}} \left ( 2g_{2} + 4g_{3} \right ) \nonumber \\
~ & + & \frac{2(1 - f)^{3}}{r^{5}} \left (4g_{4} + 2g_{5} + g_{6} \right ) \nonumber \\
~ & + & \frac{8f(1 - f)^{2}}{r^{5}} \left ( g_{7}+ g_{8} \right ) \ .
\label{eq: defnu2}
\end{eqnarray}
For $f = 1$, we get $\nu_1(r) = \nu_2(r) = -g_0 \zeta^6 r$ so that Eqs.(\ref{eq: ss1}) and (\ref{eq: ss2})
are equal hence the solution for $N_r(r)$ is the same. However, differently from what stated in TC09, this is
not the only possibility. Indeed, in order to have the same $N_r(r)$ solving both Eqs.(\ref{eq: ss1}) and
(\ref{eq: ss2}), one must have $\nu_1(r) = \nu_2(r)$ which is possible by equating the coefficients of the
terms with equal orders in $r$ in the two functions. Comparing $\nu_1(r)$ and $\nu_2(r)$, one has just to
equate the coefficients of the terms in $r^{-3}$ and $r^{-5}$ thus obtaining the following two equations\,:

\begin{equation}
(1 - f) \left [ 2 g_2 (1 + 7f) + g_3 (1 + 5f) \right ] = (1 - f)^2 \left (2 g_2 + 4 g_3 \right ) \ ,
\label{eq: sys1}
\end{equation}

\begin{eqnarray}
2 (1 - f)^2 \left [ 4 g_4 (1 + 23 f) + 2 g_5 (1 + 17 f) + g_6 (1 + 14 f) \right ] + \nonumber \\
8 f (1 - f) \left [ 2 g_7 (1 + 7 f) + g_8 (1 - 4 f) \right ] = \nonumber \\
2 (1 - f)^3 \left ( 4 g_4 + 2 g_5 + g_6 \right ) +
8 f (1 - f)^2 \left ( g_7 + g_8 \right ) \ ,
\label{eq: sys2}
\end{eqnarray}
which allows us to set two of the quantities $(g_1, \ldots, g_8, f)$ as a function of the others.
Provided this condition has been satisfied, the solution of Eq.(\ref{eq: ss1}) automatically solves
also the integral condition (\ref{eq: ss2}) so that we can limit our attention only to Eq.(\ref{eq: ss1}).
This is a linear first order equation in $N_r^2(r)$ which can be analitically solved giving\,:

\begin{equation}
N_r(r) = \pm \sqrt{\frac{{\cal{M}}}{r^2} - \frac{N^2}{2f^2} \left ( \frac{A}{3} r^2 + B
- \frac{C}{r^2} - \frac{D}{3 r^4} \right )}
\label{eq: nrsol}
\end{equation}
where ${\cal{M}}$ is an integration constant and we have defined\footnote{Note that,
following TC09, we have used units in which $Z = 1$ with $Z$ a dimensional parameter
of the HL theory. In order to have $(A, B, C, D)$ expressed in the more common $c = 1$
units, one has simply to multiply each term in Eq.(\ref{eq: defhlpar}) by $\zeta^{-4}$
with $\zeta$ having, in these units, the dimension of a length.}

\begin{equation}
\left \{
\begin{array}{lll}
\displaystyle{A} & = & \displaystyle{-g_0 \zeta^6} \\
~ & ~ & ~ \\
\displaystyle{B} & = & \displaystyle{2 \zeta^4 (1 - f)} \\
~ & ~ & ~ \\
\displaystyle{C} & = & \displaystyle{2 \zeta^2 (1 - f) \left [ 2 g_2 (1 + 7 f) + g_3 ( 1 + 5 f) \right ]} \\
~ & ~ & ~ \\
\displaystyle{D} & = & \displaystyle{2 (1 - f)^2 \left [ 4 g_4 (1 + 23 f) + 2 g_5 (1 + 17 f ) \right .}  \\
~ & ~ & ~ \\
~ & + & \displaystyle{\left . g_6 (1 + 14 f) \right ] + 8 (1 - f) f \left [ 2 g_7 (1 + 7 f) \right . } \\
~ & ~ & ~ \\
~ & + & \displaystyle{\left . g_8 (1 - 4 f) \right ]}
\end{array}
\right . \ .
\label{eq: defhlpar}
\end{equation}
In order to derive the gravitational potential, we have first to rewrite the line element
in the usual Schwartzschild\,-\,like form, i.e.

\begin{displaymath}
ds^2 = -g_{00}(r) dt^2 + g_{rr}(r) dr^2 + r^2 d\Omega^2
\end{displaymath}
and then use the general relations\,:

\begin{displaymath}
g_{00}(r) = N^2 - f N_r^2 \ \ , \ \ g_{rr}(r) = \frac{N^2}{f \left ( N^2 - f N_r^2 \right )} \ .
\end{displaymath}
Fixing $N = 1$ (which is always possible by a rescaling of the $t$ coordinate),  we then get\,:

\begin{equation}
g_{00}(r) = 1 - \frac{f {\cal{M}}}{r} + \frac{1}{2 f} \left ( \frac{A}{3} r^2 +
B - \frac{C}{r^2} - \frac{D}{3 r^4} \right ) \ ,
\label{eq: g00}
\end{equation}

\begin{equation}
g_{rr}(r) = 1/f g_{00}(r) \ .
\label{eq: grr}
\end{equation}
Since the gravitational potential generated by a pointlike mass particle may  be easily recovered from the
usual relation $g_{00}(r) = 1 + 2 \Phi(r)$, we then find\,:

\begin{equation}
\Phi(r) = -\frac{f {\cal{M}}}{2 r} + \frac{1}{4 f}
\left ( \frac{A}{3} r^2 + B - \frac{C}{r^2} - \frac{D}{3 r^4} \right ) \ .
\label{eq: prephi}
\end{equation}
Note that, up to now, the two integration constants $f$ and ${\cal{M}}$ are still undetermined.
However, in order to recover the usual Newtonian potential in the GR limit (i.e., for $A = B = C = D = 0$),
we must set $f {\cal{M}} = 2 G m$ with $m$ the mass of the gravitational field source. It is worth noting
that, for $m = 0$, the potential (\ref{eq: prephi}) leads to unphysical divergences in the metric because
of the terms in $C/r^{-2}$ and $D/3 r^{4}$. In order to avoid this problem, we must impose that both $C$ and
$D$ vanish for $m = 0$. The first trivial choice is to impose $C = D = 0$ identically so that one recovers
the usual Schwartzschild\,-\,de Sitter case. As a more attractive possibility, one can postulate a dependence
of the metric coefficient $f$ on the mass. In such a case, Eqs.(\ref{eq: sys1}) and (\ref{eq: sys2}) can not be
read as a relation among $f$ and the HL couplings $g_i$ since this will induce a dependence of the HL Lagrangian
on the mass which is not possible. As such, we consider them as algebraic equations for $f$ so that, in order to
be fulfilled, we must equate the terms with equal powers of $f$ on the two sides. We then obtain\,:

\begin{displaymath}
\left \{
\begin{array}{l}
\displaystyle{g_2 = g_3 = 0} \\
~ \\
\displaystyle{g_4 = -(36 g_5 + 15 g_6 + 4 g_7)/96} \\
~ \\
\displaystyle{g_8 = (16/3) g_7} \\
\end{array}
\right .
\end{displaymath}
which implies $C = 0$ and

\begin{displaymath}
D = \frac{(1 - f)^2}{12} \left [ 12 (1 - f) g_5 - 9 (1 - f) g_6 + 4 (1 - 153 f) g_7 \right ] \ .
\end{displaymath}
With these conditions, we can finally write the gravitational potential generated by a points mass $m$ as\,:

\begin{equation}
\Phi(r) = \Phi_N(r) + \Phi_{HL}(r)
\label{eq: phipoint}
\end{equation}
where $\Phi_N(r) = - Gm/r$ is the Newtonian potential and

\begin{equation}
\Phi_{HL}(r)=\frac{A}{12f}r^{2}+\frac{B}{4f}-\frac{D}{12f}\frac{1}{r^{4}}
\label{eq:HL1point}
\end{equation}
is the correction due to HL theory. Is is worth noting that, while the second term is simply a constant
having no impact on the dynamics, the other two terms have a simple relation with previous results in literature.
Indeed, the first one plays the role of a cosmological constant thus remembering the Schwartzschild\,-\,de Sitter
solution already found in TC09. On the other hand, the third term has the same asymptotical behaviour of the corrections
to the Newtonian potential in the Kehagias\,-\,Sfestos (2009, hereafter KS) static spherically\,-\,symmetric solution of
HL gravity. Iorio \& Ruggiero (2009) have shown that such corrections are proportional to $m^{2}r^{-4}$ thus suggesting
that $D$ should actually be a function of the mass $m$ generating the gravitational field, a point which we will come back to later.

The corrective term may be conveniently written as\,:

\begin{equation}
\Phi_{HL}(r) = - \frac{G M_{\odot}}{r_s} \left [
- \left ( \frac{\eta}{\eta_A} \right )^2 - \frac{B r_s}{4 f G M_{\odot}}
+ \left ( \frac{\eta}{\eta_D} \right )^{-4} \right ]
\label{eq: phihlpoint}
\end{equation}
with $M_{\odot}$ the Sun mass and $r_s$ an arbitrary chosen reference radius introduced to
define the dimensionless quantity $\eta = r/r_s$. In Eq.(\ref{eq: phihlpoint}), we have
finally defined the scaling radii\,:

\begin{equation}
\left \{
\begin{array}{lll}
\displaystyle{\eta_A} & = & \displaystyle{\left ( \frac{12 f G M_{\odot}}{A r_s} \right )^{1/2}} \\
~ & ~ & ~ \\
\displaystyle{\eta_D} & = & \displaystyle{\left ( \frac{D r_s^{-3}}{12 f G M_{\odot}} \right )^{1/4}} \\
\end{array}
\right .
\label{eq: defradii}
\end{equation}
which can be used as model parameters instead of the $(A, D)$ coefficients. Some caveats are in
order here, First, note that, in order to preserve the interpretation of $(r_A, r_D) = r_s \times (\eta_A, \eta_D)$  as physical radii, one has to postulate that the $(A, D)$ parameters are positive quantities thus narrowing the space of the parameters $(g_0, g_5, g_6, g_7, f)$ entering $(A, D)$. Should $A$ or $D$ be negative, we could nonetheless
define $(r_A, r_D)$ as above and accordingly change the sign of the corresponding term in the potential.
For definiteness, we will both $A$ and $D$ are positive so that the potential is given by Eq.(\ref{eq: phihlpoint})
without any sign change. We then remember that, in our scheme, $f$ is a dimensionless function of $\mu = m/M_{\odot}$, but its functional dependence can not be obtained in any way. The only constraint we have is $f(\mu = 0) = 1$
so that $D = 0$ and we recover the Schwartzschild\,-\,de Sitter solution in accordance with the fact that $g_0$
(and hence $A$) play the role of a cosmological constant term. In order to parametrize our ignorance we can
redefine the above scaling radii as\,:

\begin{equation}
\left \{
\begin{array}{lll}
\displaystyle{\eta_A} & = & \displaystyle{\left ( \frac{12 f_{\odot} G M_{\odot}}{A r_s} \right )^{1/2}}
\left [ \frac{f(\mu)}{f_{\odot}} \right ]^{1/2} \\
~ & ~ & ~ \\
\displaystyle{\eta_D} & = & \displaystyle{\left ( \frac{D_{\odot} r_s^{-3}}{12 f_{\odot} G M_{\odot}} \right )^{1/4}}
\left [ \frac{D(\mu)}{D_{\odot}} \frac{f_{\odot}}{f_(\mu)} \right ]^{1/4} \\
\end{array}
\right .
\label{eq: defradiibis}
\end{equation}
where quantities labelled with a $\odot$ are evaluated for $\mu = 1$. It is worth wondering whether some hint
on the functional expression of $f(\mu)$ can be retrieved. Up to now, we have postulated $f = f(\mu)$ in order to
get a mathematically viable solution other than the Schwartzschild\,-\,de Sitter one. It is, however, worth noticing
that a dependence of $f$ on the mass may also be physically motivated. Although the HL theory is obtained by modifying 
General Relativity, it is still true that the properties of the spacetime are determined by the source of the gravitational 
field. As such, one can expect that the corrective HL term to the Newtonian potential is still related
to the only property characterizing the field, i.e. the source mass $m$. Since the $(A, B, D)$ coefficients in Eq.(\ref{eq: hl1point}) 
are related to the HL couplings (and hence are universal quantities), the only way to introduce a dependence of the solution on the source 
properties is to postulate $f = f(\mu)$. Although qualitative, this discussion shows that our mathematical assumption is actually deeply 
related to a physical motivation.

Finally, we note that the second term in Eq.(\ref{eq: phihlpoint}) simply adds a constant to the potential
which has no effect in any situation of interest so that we will henceforth neglect this term. Note that this by
no way means that $B$ can be set to zero. Indeed, $B = 0$ means $f = 1$ so that also $D$ vanish and
we go back to the Schwartzschild\,-\,de Sitter solution, while we are here interested in the more general case.
We therefore assume $B \neq 0$, but nevertheless neglect its contribution hereafter because drops off from
the derivation of the quantities we are interested in.

\section{The rotation curve}\label{The rotation curve}

The modified gravitational potential derived above deviates from the Newtonian one because of
the additive terms in Eq.(\ref{eq: phihlpoint}). Depending on the values of the scaling radii
$(r_A, r_D)$, we can have different situations. As a general remark, we note that, for $\eta
<< 1$, the last term in Eq.(\ref{eq: phihlpoint}) increases the potential with respect to the
Newtonian one. On the contrary, for $\eta >> 1$, the first term boosts the potential making it
deviating from the Keplerian fall off. When considering the rotation curve, $v_c(r) = r d\Phi/dr$,
we thus get a circular velocity which may significantly differ from the Newtonian one being
larger than the classical value both in the inner and outer regions. It is therefore worth
wondering whether such deviations may help in fitting spiral galaxies rotation curves without
the need of other mass than the visible one. That is to say, we are here interested in investigating
whether the HL modifications to the potential can also play the role of an effective dark halo. As
a preliminary remark, we stress that a similar analysis could in principle be made also for the
KS solution. However, in that case, the corrections fade away as $r^{-4}$ and there is no cosmological
constant term. It is therefore expected that the correction to the rotation curve in the outer regions
are negligible so that we argue that the KS solution can not play the role of an effective dark halo.

To this end, we have first to derive an expression for the rotation curve of an extended system
generalizing the procedure adopted in the Newtonian gravity framework. In that case, the circular
velocity in the equatorial plane is given by $v_c^2(R) = R d\Phi/dR|_{z = 0}$, with $\Phi$ the total
gravitational potential. Thanks to the superposition principle and the linearity of the point mass
potential on the mass $m$, this latter is computed by adding the contribution from infinitesimally small
mass elements and then transforming the sum into an integral over the mass distribution. Such a simple
procedure can not be applied in the HL case since the corrective term $\Phi_{HL}$ does depend on $m$ in
a way that we do not explicitly know so that a nonlinear dependence can not be excluded a priori. In order
to overcome this difficulty, we have developed an alternative procedure which can be actually applied to any
kind of potential provided some general conditions hold.

As a starting point, let us denote by $F_{p}(m, r)$ the gravitational force (per unit of test mass
particle) generated by a point mass $m$. Whatever is the dependence of $F_p$ on $m$, it is always
true that the total force due to $N$ particles of mass $m$ is the sum of the single forces. As such,
taking the continuum limit, we can estimate the (magnitude of the) total force as\,:

\begin{equation}
F({\bf r}) = \int{n(m, {\bf r'}) F_p(m, |{\bf r} - {\bf r'}|) dm dV}
\label{eq: forcetot}
\end{equation}
where the integral is over the full mass range and volume and $n(m, {\bf r})$ is the star mass
function (hereafter $MF$), i.e. the number of stars in the volume element $dV$ with mass between
$m$ and $m + dm$. Because of its definition, we have\,:

\begin{equation}
\int_{m_{min}}^{m_{max}}{n(m, {\bf r}) m dm} = \rho({\bf r})
\label{eq: mfnorm}
\end{equation}
with $\rho({\bf r})$ the mass density. As usual in literature, we will adopt the factorization
hypothesis thus writing $n(m, {\bf r}) = \psi(m) \tilde{\rho}({\bf r})$ with $\psi(m)$ the
local\footnote{Note that the term {\it local} refers to the Solar neighborhood when the galaxy is
the Milky Way. For external galaxies, by {\it local}, we mean the MF in the neighborhood of a
suitably chosen reference radius $R_0$.} MF and $\tilde{\rho} = \rho({\bf r})/\rho_0$ with $\rho_0
= \rho({\bf R_0})$. Defining $\mu = m/M_{\odot}$, we have the following normalization condition for
the local MF\,:

\begin{displaymath}
\int_{\mu_{min}}^{\mu_{max}}{\mu \psi(\mu) d\mu} = \rho_0/M_{\odot}^2
\end{displaymath}
so that henceforth we use as local MF the quantity ${\cal{N}}_1 \psi(\mu)$
with

\begin{equation}
{\cal{N}}_1 = \frac{\rho_0}{M_{\odot}^2} \left [ \int_{\mu_{min}}^{\mu_{max}}{\mu \psi(\mu) d\mu} \right ]^{-1} \ .
\label{eq: defn1}
\end{equation}
Let us now assume that the point mass gravitational force may be factorized as\,:

\begin{displaymath}
F_p(\mu, r) = \frac{G M_{\odot}}{r_s^2} f_{\mu}(\mu) f_{r}(\eta)
\end{displaymath}
with $f_{\mu}$ and $f_r(r)$ dimensionless functions depending on the particular form of the
point mass gravitational potential $\Phi_p$. Remembering that ${\bf F}_p = - \nabla \Phi_p$,
it is only a matter of algebra to show that it is\,:

\begin{displaymath}
f_{\mu}(\mu) = \mu \ \ , \ \ f_{r}(\eta) = 1/\eta^2 \ \ ,
\end{displaymath}
for the Newtonian potential. For the HL corrective potential, we can split it as the sum of two terms, i.e.

\begin{displaymath}
\Phi_{HL}(r, r_A, r_D) =  \Phi_{A}(r, r_A) + \Phi_D(r, r_D) \ ,
\end{displaymath}
and then obtain\,:

\begin{displaymath}
f_{r} = \left ( \frac{2}{\eta_{A \odot}} \right ) \left ( \frac{\eta}{\eta_{A \odot}} \right ) \ \ ,
\ \ f_\mu = \left [ \frac{f(\mu)}{f_{\odot}} \right ]^{-1} \ \ ,
\end{displaymath}
for $\Phi_A$ and

\begin{displaymath}
f_{r} = \left ( \frac{4}{\eta_{D \odot}} \right ) \left ( \frac{\eta}{\eta_{D \odot}} \right )^{-5} \ \ , \ \
f_\mu = \frac{D(\mu)}{D_{\odot}} \frac{f_{\odot}}{f(\mu)} \ \ ,
\end{displaymath}
for the $\Phi_D$ term. We now use cylindrical coordinates $(R, \theta, z)$ and the
corresponding dimensionless variables $(\eta, \theta, \zeta)$ (with $\zeta = z/r_s$) and rely on the factorization hypotheses
for both the MF and the point mass force to finally get\,:

\begin{eqnarray}
F({\bf r}) & = & G \ \rho_0 \ r_s \nonumber \\
~ & \times & \frac{\int_{\mu_{min}}^{\mu_{max}}{f_{\mu}(\mu) \psi(\mu) d\mu}}
{\int_{\mu_{min}}^{\mu_{max}}{\mu \psi(\mu) d\mu}} \nonumber \\
~ & \times & \int_{0}^{\infty}{\eta' d\eta' \int_{-\infty}^{\infty}{d\zeta'
\int_{0}^{\pi}{f_{r}(\Delta) \tilde{\rho}(\eta', \theta', \zeta') d\theta'}}}
\label{eq: forcetotgen}
\end{eqnarray}
with the shorthand notation

\begin{equation}
\Delta = \left [ \eta^2 + \eta'^2 - 2 \eta \eta' \cos{(\theta - \theta')} + (\zeta - \zeta')^2 \right ]^{1/2} \ .
\label{eq: defdelta}
\end{equation}
The circular velocity in the equatorial plane along the major axis (which is the quantity typically measured for
spiral galaxies) will be simply $v_c^2(R) = R F(R, \theta = z = 0)$. Since we will be interested in axisymmetric
systems, we can set $\tilde{\rho} = \tilde{\rho}(\eta, \zeta)$. Moreover, our systems will be spiral galaxies, hence
made out of a spheroidal bulge and a circular disk, so that a convenient choice for the scaling radius $r_s$ will be
the disk scalelenght $R_d$. Under these assumptions, the rotation curve may then be evaluated as\,:

\begin{eqnarray}
v_c^2(R) & = & G \ \rho_0 \ R_d^2 \ \eta \nonumber \\
~ & \times & \frac{\int_{\mu_{min}}^{\mu_{max}}{f_{\mu}(\mu) \psi(\mu) d\mu}}
{\int_{\mu_{min}}^{\mu_{max}}{\mu \psi(\mu) d\mu}} \nonumber \\
~ & \times & \int_{0}^{\infty}{\eta' d\eta' \int_{-\infty}^{\infty}{\tilde{\rho}(\eta', \zeta') d\zeta'
\int_{0}^{\pi}{f_{r}(\Delta_0) d\theta'}}}
\label{eq: rotcurvegen}
\end{eqnarray}
with

\begin{equation}
\Delta_0 = \Delta(\theta = \zeta = 0) =
\left [ \eta^2 + \eta'^2 - 2 \eta \eta' \cos{\theta'} + \zeta'^2 \right ]^{1/2} \ .
\label{eq: defdeltazero}
\end{equation}
It is worth noting that, except in very particular cases (e.g., the Newtonian potential
for a spherically symmetric mass distribution), the integrals over the coordinates in
Eq.(\ref{eq: rotcurvegen}) have to be evaluated numerically. For computational reasons, it
is useful to exchange the order of integration and resort to the logarithmic variables
$\lambda = \log{\eta}$ and $\omega = \log{\zeta}$ so that we get the following equivalent
expression for the rotation curve\,:

\begin{eqnarray}
v_c^2(R) & = & G \ \rho_0 \ R_d^2 \ (\ln{10})^2 \ {\rm dex}(\lambda) \nonumber \\
~ & \times & \frac{\int_{\mu_{min}}^{\mu_{max}}{f_{\mu}(\mu) \psi(\mu) d\mu}}
{\int_{\mu_{min}}^{\mu_{max}}{\mu \psi(\mu) d\mu}} \nonumber \\
~ & \times & \int_{-1}^{1}{\frac{\tilde{{\cal{R}}}(\lambda, \xi)}{\sqrt{1 - \xi^2}} d\xi}
\label{eq: rotcurvenum}
\end{eqnarray}
with ${\rm dex}(x) = 10^{x}$ and

\begin{equation}
{\tilde{R}} = \int_{-\infty}^{\infty}{{\rm dex}(2 \lambda') d\lambda'
\int_{-\infty}^{\infty}{f_{r}(\Delta_0) \tilde{\rho}(\lambda', \omega')
{\rm dex}(\omega') d\omega'}} \ .
\label{eq: defrtilde}
\end{equation}
Eqs.(\ref{eq: rotcurvegen}) and (\ref{eq: rotcurvenum}) are fully general and
can be used to compute the rotation curve provided the expression for $f_{\mu}(\mu)$
and $f_r(\eta)$ are given. As a consistency check, it is easy to show that, for the
Newtonian potential, the term depending on the MF is identically unity so that
Eq.(\ref{eq: rotcurvegen}) reduces to a simple rewriting of the standard result. For
the HL term, we get a dependence on the MF through the multiplicative term on the
second row of Eq.(\ref{eq: rotcurvegen}). It is worth stressing that the MF here only
plays the role of scaling up or down the rotation curve. This is a consequence of the
HL corrective term (\ref{eq: phihlpoint}) not depending on $m$. As such, indeed, one has
simply to sum the contribution of all the stars notwithstanding their mass and this is
indeed what the MF term gives in the HL case. The total rotation curve for the HL theory
will finally be given as\,:

\begin{displaymath}
v_c^2(R) = v_N^2(R) + v_{HL}^2(R) = v_N^2(R) + v_A^2(R) + v_D^2(R)
\end{displaymath}
where one can use the literature results for the computation of the Newtonian term $v_N^2(R)$
and our general rule with $f_{\mu} = 1$ and $f_{r}(\eta)$ given above to estimate the HL
contribution $v_{HL}^2(R)$. Note that, in order to evaluate $v_c^2(R)$, we need to know not only
the mass density (as in the Newtonian only case), but also the local MF because of the HL term.

\section{The Milky Way rotation curve}

The HL theory has been originally conceived as an attempt to solve some
quantum gravity related problems, but has soon attracted a lot of interest
even as an alternative model for dark energy because of the presence of a
cosmological constant like term. We have here shown that also the gravitational
potential is modified with respect to the standard Newtonian one so that it is
worth wondering if the additional terms may help in reconciling the data on the
rotation curves with what is predicted from the visible matter only. To this end,
one should fit the rotation curves of many spiral galaxies and find out that the
fit is indeed satisfactorily good and the model parameters $(r_A, r_D)$ are
the same for all the galaxies being related to the HL Lagrangian couplings and hence
universal quantities. Such a task is actually quite complicated because of the need
for the knowledge of the local MF which is not at all constrained in external galaxies.
Moreover, what one observes for these systems is the surface brightness profile due to the
two visible components, namely the bulge and the disk. What we need to evaluate the
rotation curve is, however, the mass density so that, even assuming simple functional
forms for the bulge and the disk, we still do not have any knowledge of the bulge and
disk mass\,-\,to\,-\,light ratios thus adding more parameters (and hence severe degeneracies)
to be determined. As a first preliminary test, we therefore limit our attention to the
Milky Way (hereafter MW) only. Our position within it allows us to determine the local MF
(e.g., by star counts or converting the observed luminosity function into a MF through an
empirically determined $M/L$ ratio) thus reducing the uncertainties of the problem. Moreover,
we also have direct determinations of the galactic parameters of interest so that the only
unknown quantities are the HL parameters $(r_A, r_D)$ thus strongly reducing the
possibility of degeneracies. In the following, we first describe the MW mass models and the
data on the rotation curve and then present the fitting procedure and the results.

\subsection{The mass models}

It is common to describe a spiral galaxies as the sum of a visible component (made out of
stars and gas) embedded in a dark halo (mainly populated by cold dark matter particles). Lacking
any definitive laboratory evidence for DM, the only reason why one has to include it in
galaxy modelling is to fit the rotation curves data. Since this evidence implicitly assumes
that the Newtonian potential theory is the correct one, there is actually no compelling reason
why one should add a priori a dark halo to the visible components in a modified gravity framework
as the current HL theory. We therefore model the MW as made out of vibile matter only distributed
in a spheroidal bulge\footnote{Actually, there are different evidences that the bulge has triaxial
structure. We nevertheless use a less detailed spheroidal model since the bulge contribute to the dynamics
over the range probed by the data is much smaller than the disc one, independently on the gravitational
theory adopted. Such a simplification allows us to use Eq.(\ref{eq: rotcurvegen}) without introducing any
significative bias.} and a thick disc.

We follow Dehnen \& Binney (1998, hereafter DB98) describing the bulge as a truncated power\,-\,law model,
i.e. the scaled mass density reads\,:

\begin{equation}
\tilde{\rho}_b(R, z) = \left ( \frac{\varrho}{R_b} \right )^{-\gamma} \left (1 + \frac{\varrho}{R_b}
\right )^{\gamma - \beta} \exp{\left ( - \frac{\varrho^2}{R_t^2} \right )}
\label{eq: rhobulge}
\end{equation}
with $\varrho^2 = R^2 + z^2/q^2$. Fitting the model to the infared photometric $COBE/DIRBE$ data
yields values for the model parameters, namely\,:

\begin{displaymath}
\beta = \gamma = 1.8 \ \ , \ \ q = 0.6 \ \ ,
\ \ R_b = 1 \ {\rm kpc} \ \ ,  \ \ R_t = 1.9 \ {\rm kpc} \ \ .
\end{displaymath}
The reference density is not determined from the photometry, but may be related
to the total bulge mass $M_b$ as\,:

\begin{equation}
\rho_{0,b} = \frac{M_b}{2 \pi \int_{0}^{\infty}{R dR \int_{-\infty}^{\infty}{\tilde{\rho}(R, z) dz}}} \ .
\label{eq: rhozbulge}
\end{equation}
Dwek et al. (1995) has found $M_b = (1.3 \pm 0.3) \times 10^{10} \ {\rm M_{\odot}}$ for the
bulge mass so that we set $M_b$ to its central value neglecting the measurement uncertainty. We have
then to set the local MF for the bulge. Zoccali et al. (2000) have determined it from the luminosity
function of lower main sequence stars finding\,:

\begin{displaymath}
\psi_b(\mu) \propto \mu^{-\beta}
\end{displaymath}
with $\beta = -1.33 \pm 0.07$ for stars in the mass range $(0.15, 1.0) \ {\rm M_{\odot}}$.
We extend it to the brown dwarfs region and neglect the measurement uncertainty thus
setting $(\beta, \mu_{min}, \mu_{max}) = (-1.33, 0.03, 1.0)$ as our bulge MF parameters.

While important in the inner regions, the bulge only plays a minor role in determining
the circular velocity over the regions probed by the data. The dynamics is here dominated
by the disc component that we model as a double exponential. The mass density then reads\,:

\begin{equation}
\tilde{\rho}_d(R, z) = \exp{\left ( - \frac{R}{R_d} \right )} \exp{\left ( -
\frac{|z|}{z_d} \right )}
\label{eq: rhodisk}
\end{equation}
where we follow DB98 fixing\,:

\begin{displaymath}
R_d = \kappa_d R_0 \ \ , \ \ z_d = 0.18 \ {\rm kpc} \ \ ,
\end{displaymath}
with $\kappa_d = 0.30 \pm 0.05$ a scaling parameter and $R_0$ the Sun
distance from the MW centre. We then follow Cardone \& Sereno (2005) adopting
$\kappa_d = 0.30$ and $R_0 = 8.5 \ {\rm kpc}$ as fiducial parameters. The disk
reference density is related to the disk total mass as\,:

\begin{equation}
\rho_{0,d} = \frac{M_d}{4 \pi R_d^2 z_d}
\label{eq: rhozdisk}
\end{equation}
while the disk mass is estimated as

\begin{equation}
M_d = 2 \pi R_d^2 \Sigma_{\odot} \exp{(R_0/R_d)}
\label{eq: discmass}
\end{equation}
with $\Sigma_{\odot} = 48 \pm 8 \ {\rm M_{\odot}/pc^2}$ \cite{KG89} the disc
surface density at the Sun position. A caveat is in order here. The measured
value of $\Sigma_{\odot}$ refers to the total mass, both stars and gas. In
principle, we should separate the two components computing the total disc mass
using $\Sigma_{\star}$ instead of $\Sigma_{\odot}$, with $\Sigma_{\star} =
\Sigma_{\odot} - \Sigma_{ISM}$ and $\Sigma_{ISM} = 14.5 \ {\rm M_{\odot}/pc^2}$
\cite{OM01} the gas surface density. We should then add a further disc like
component for the gas with its density profile. As a reasonable approximation, we
can, however, assume that the gas follows the same density profile as the stellar
disc with the same scalelength radius so that the total disc mass is indeed
given by Eq.(\ref{eq: discmass}).

As a final ingredient, we need the local disc MF. This is determined quite accurately
thanks to our position in the disc plane. We refer the reader to Chabrier (2003) and
references therein for a detailed discussion of this issue motivating our choice of
adopting the Kroupa MF, i.e. a power\,-\,law MF with slope $\beta$ changing with the
mass range as\,:

\begin{displaymath}
\beta = \left \{
\begin{array}{ll}
0.3 & 0.01 \le \mu \le 0.08 \\
~ & ~ \\
1.3 & 0.08 \le \mu \le 0.50 \\
~ & ~ \\
2.3 & 0.50 \le \mu \le 1.0 \\
~ & ~ \\
4.5 & 1.0 \le \mu \le 120.0 \\
\end{array}
\right . \ .
\end{displaymath}
The four different range contributes to the local mass density according
to the following percentages\,:

\begin{displaymath}
(f_1, f_2, f_3, f_4) = (7.39, 48.21, 29.22, 15.18)\%
\end{displaymath}
so that we can compute the multiplicative term entering the rotation curve
for the HL term.

\subsection{The data}

Although not ideal targets for our test because of the poor knowledge of their
parameter and MF, external galaxies are better suited for the measurement of precise
and extended rotation curves. On the contrary, our position in the MW disc equatorial
plane makes it difficult to observationally determine this quantity. Indeed, one has
to characterize the full line of sight velocity distribution in order to correct the
observed velocity for asymmetric drift and projection effects. Moreover, the distance
to the tracer should be known with great accuracy not to bias in a dangerous way the
estimated circular velocity. Notwithstanding these difficulties, the rotation curve in
the outer regions have been measured relying on Cepheids and HII regions. We follow DB98
to estimate the circular velocity from the Cepheids data of Pont et al. (1997) and the
HII molecular clouds sample of Brand \& Blitz (1993) using their same selection criteria.
This dataset probes the radial range $8.2 \le R \ {\rm (kpc)} \ \le 18.94$ and is affected
by a large scatter due to both measurement errors and the inhomogeneity of the data. In order
to smooth the data without introducing any bias or spurious correlations, we use the local
regression method \cite{L99}. Originally proposed by Cleveland (1979) and further developed by Cleveland
and Devlin (1988), the local regression technique combines much of the simplicity of linear least
squares regression with the flexibility of nonlinear regression. The basic idea relies on fitting
simple models to localized subsets of the data to build up a function that describes the
deterministic part of the variation in the data, point by point. Actually, one is not required to
specify a global function of any form to fit a model to the data so that there is no ambiguity in
the choice of the interpolating function. Indeed, at each point, a low degree polynomial is fit to
a subset of the data containing only those points which are nearest to the point whose response is
being estimated. The polynomial is fit using weighted least squares with a weight function which
quickly decreases with the distance from the point where the model has to be recovered. We hence use
this method\footnote{See, e.g., Capozziello et al. 2007 for a step\,-\,by\,-\,step description.} to
smooth the sample points and cut data with $R > 14 \ {\rm kpc}$ since the local regression method
becomes unreliable for these points because of the sparseness of the sample in this region. We finally
give it away points with $S/N \le (S/N)_{min}$ where the threshold signal\,-\,to\,-\,noise ratio
has been set to $(S/N)_{min} \simeq 8$ for reasons explained below. We will refer to this sample as
the DB data in the following.

Another possible tracer of the velocity field in the outer regions is represented by the Blue Horizontal
Branch (BHB) stars. Photometry and spectra from the Sloan Digital Sky Survey (SDSS) for a sample of 2466
BHB stars allow to determine the rotation curve up to $\sim 60 \ {\rm kpc}$ as described in Xue et al.
(2008, hereafter X08). In order to extract $v_c(R)$ taking care of all the possible projection effects and
systematic biases (due to, e.g., the survey mask and targetting algorithm), X08 have relied on mock matched
observations based on two different cosmological simulations of MW\,-\,like galaxies. As such, there are two
different datasets labelled as $N$ and $S$ depending on the simulation adopted. We here use only the $N$
dataset since it is in better agreement with the DB sample over the range of overlap. Note that this sample
(referred to hereafter as the SDSS dataset) is made out of only 10 points obtained by binning the 2466 BHB
stars in almost equally spaced radial bins covering the range $(7.5, 55) \ {\rm kpc}$. The S/N is quite high
with a median value $(S/N)_{med} \simeq 8$ so that we set $(S/N)_{min} = (S/N)_{med}$ in order to give
the same statistical weight to both datasets. As a final remark, we note that, while the DB data probe
only the outer disc being $R/R_d \simeq 3.2 - 5.4$, the SDSS sample extends mainly in the halo region
most of the data having $R/R_d > 4$ and extending up to $R/R_d \simeq 20$. As such, the two sample nicely
complement each other and allow us to check whether the outer rotation curve may be reproduced without any
dark matter contribution and, at the same time, still preserving the agreement with the data in the disc
region.

\subsection{Fitting procedure}

In order to constrain the HL model parameters, we employ a standard Bayesian approach
first defining the likelihood function as\,:

\begin{eqnarray}
{\cal{L}}({\bf p}) & \propto & \exp{\left [ - \frac{\chi^2({\bf p})}{2} \right ]} \nonumber \\
~ & = & \exp{\left \{ - \frac{1}{2} \sum{\left [ \frac{v_c^{obs}(R_i) - v_c^{th}(R_i)}{\varepsilon_i}
\right ]^2} \right \}}
\label{eq: deflike}
\end{eqnarray}
where ${\bf p} = (\log{\eta_A}, \log{\eta_C}, \log{\eta_D})$ are he HL model parameters\footnote{We use
logarithmic units in order to explore a wider range. Note also that, rigorously speaking, Eq.(\ref{eq:
deflike}) is not correct since the SDSS data points are somewhat correlated being obtained by a binning
procedure. However, since the bin spacing is quite large, it is likely that the correlation matrix (not
available) is close to diagonal so that Eq.(\ref{eq: deflike}) is essentially correct.}, $v_c^{obs}(R_i)$
and $v_c^{th}(R_i)$ are the observed (with a measurement error $\varepsilon_i$) and theoretically predicted
values of the circular velocity at the radius $R_i$ of the i\,-\,th point and the sum runs over the 101 DB
and 10 SDSS data points.

The best fit is obtained by maximizing the likelihood ${\cal{L}}({\bf p})$, but it is worth stressing that,
according to the Bayesian philosophy, the best estimate of the parameter $p_i$ is not the best fit one. On the
contrary, one has to marginalize over the remaining parameters and look at the shape of the marginalized
likelihood function defined as\,:

\begin{displaymath}
{\cal{L}}_i(p_i) \propto \int{{\cal{L}}({\bf p}) dp_1 \ldots dp_{i-1} dp_{i+1} \ldots dp_n}
\end{displaymath}
with $n$ the total number of parameters. Actually, what we do is running a Monte Carlo Markov Chain
code to efficiently explore the three dimensional parameter space $(\log{\eta_A}, \log{\eta_C},
\log{\eta_D})$ and use of the histogram of the values for the parameter $p_i$ to estimate the mean,
the median and the $68$ and $95\%$ confidence ranges. Note that, because of degeneracies among the
model parameters, the best fit parameters ${\bf p}_{bf}$ may also differ from the maximum likelihood
ones ${\bf p}_{ML}$, i.e. the set obtained by maximizing each of the marginalized likelihood functions.

\subsection{Results}

As a preliminary discussion, it is worth clearly stating which are the parameters we can
constrain. Eq.(\ref{eq: rotcurvegen} shows that, in order to compute the HL contributions to
the rotation curve, namely $v_A^2(R)$ and $v_D^2(R)$, we must know the functional expression
of $f(\mu)$ entering $f_\mu(\mu)$ through the scaling radii $(r_A, r_D)$. Since we do not know
this function, we can not separately constrain the parameters $(r_{A \odot}, r_{D \odot})$. It is, however,
easy to show that\,:

\begin{displaymath}
\kappa_A v_A^2(R, \eta_{A, \odot}) = v_A^2(R, \eta_{A1}) \ ,
\end{displaymath}

\begin{displaymath}
\kappa_D v_D^2(R, \eta_{D, \odot}) = v_D^2(R, \eta_{D1}) \ ,
\end{displaymath}
where we have defined\,:

\begin{displaymath}
\kappa_A = \frac{\int_{\mu_{min}}^{\mu_{max}}{\left [ f(\mu)/f_{\odot} \right ]^{-1} \psi(\mu) d\mu}}
{\int_{\mu_{min}}^{\mu_{max}}{\psi(\mu) \mu d\mu}} \ ,
\end{displaymath}

\begin{displaymath}
\kappa_D = \frac{\int_{\mu_{min}}^{\mu_{max}}{\left [ D(\mu)/D_{\odot} \right ] \left [ f(\mu)/f_{\odot} \right ]^{-1} \psi(\mu) d\mu}}
{\int_{\mu_{min}}^{\mu_{max}}{\psi(\mu) \mu d\mu}} \ ,
\end{displaymath}
while $\eta_A = r_A/R_{eff}$ and $\eta_D = r_D/R_{eff}$ are considered as constant parameter not depending
on $\mu$ and we assume that $v_A^2(R, \eta_A)$ and $v_D^2(R, \eta_D)$ are evaluated setting $f_{\mu} = \mu$ in
Eq.(\ref{eq: rotcurvegen}). Our fit will then give constraints on $(\eta_A, \eta_D)$, while those
on $(\eta_{A \odot}, \eta_{D \odot})$ could be derived provided a theoretically motivated functional expression
for $f(\mu)$ is given.

\begin{figure}
\includegraphics[width=8.5cm]{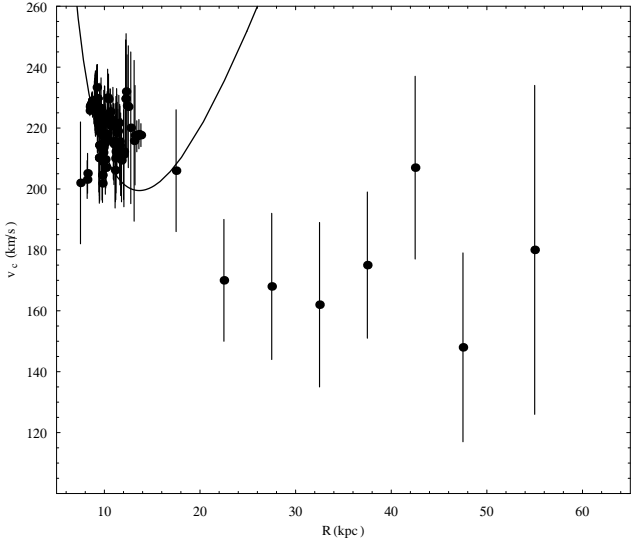}
\caption{Best fit HL model rotation curve superimposed to the DB and SDSS data.}
\label{fig: bfplot}
\end{figure}

Skipping to logarithmic units to investigate a larger range, we run a single chain with 150000 points reduced
to $\sim 10000$ after cutting out the initial burn in phase and thinning the chain (with a step of 10) to avoid
spurious correlations. The best fit point turns out to be\,:

\begin{displaymath}
(\log{\eta_A}, \log{\eta_D}) = (1.051, -0.060)
\end{displaymath}
giving $\chi^2/d.o.f. = 523.56/109 = 4.80$. Needless to say, such a high reduced $\chi^2$ is a strong
evidence against the HL model with no dark matter. A simple look at Fig.\,\ref{fig: bfplot} shows that
the large $\chi^2$ is due mainly to the SDSS data. Indeed, it is $\chi^2_{DB}/d.o.f. = 235.91/99 = 2.38$
and $\chi^2_{SDSS}/d.o.f. = 287.65/8 = 35.96$ for the DB and SDSS samples, respectively. Although the model
is clearly at odds with the data, we give below, for completeness, the constraints on the parameters from
the analysis of their histograms along the chain. For $\log{\eta_A}$ we get\,:

\begin{displaymath}
\langle \log{\eta_A} \rangle = 1.049 \ \ , \ \
(\log{\eta_A})_{med} = 1.050 \ \ , \ \
\end{displaymath}

\begin{displaymath}
68\% : (1.040, 1.059) \ \ , \ \
95\% : (1.028, 1.066) \ \ , \ \ ,
\end{displaymath}
while it is\,:

\begin{displaymath}
\langle \log{\eta_D} \rangle = -0.061 \ \ , \ \
(\log{\eta_D})_{med} = -0.060 \ \ , \ \
\end{displaymath}

\begin{displaymath}
68\% : (-0.064, -0.056) \ \ , \ \
95\% : (-0.068, -0.054) \ \ , \ \
\end{displaymath}
for $\log{\eta_D}$. Converting median values in linear units, we get\,:

\begin{displaymath}
(r_A, r_D) = (11.2, 0.87) \ {\rm kpc}
\end{displaymath}
which allows us to make some qualitative interpretation of the results.
Considering the form of the potential for the point mass case, we see
that only the first term can actually boost the stellar contribution to the
rotation curve. This term essentially translates in a contribution to $v_c(R)$ which
linearly increases with $R/r_A$ so that $r_A$ must be small in order to fit the data
in the DB probed region without dark matter. On the other hand, a too small value may
lead to a $v_c(R)$ soon diverging thus spoiling down the agreement with the SDSS data.
The value $r_A = 11.2 \ {\rm kpc}$ we find is the result of this compromise, but it is
nevertheless unable to reconcile the model with the SDSS data. A similar compromise drives
the fit in the estimates of $\log{\eta_D}$. Because of the scaling with $r^{-4}$
in the point mass potential, the term entering $r_D$ may contribute
to the outer region circular velocity only if $r_D$ is quite large. But, in such a case,
$R/r_D$ will be much smaller than 1 in the inner regions thus making the contribution
of this term to overcome by orders of magnitude the Newtonian one in the regions where this
latter is yet able to fit the data. As a consequence, $r_D$ must be small so that it does
not contribute enough to boost $v_c$ with respect to the Newtonian value hence motivating the
need for a not too large $r_A$. Indeed, we find an almost linear correlation between $\log{\eta_A}$
and $\log{\eta_D}$ along the chain motivated by the fact that the larger is $r_D$, the larger is the
contribution to $v_c(R)$ of the $r^{-4}$ term of the potential and hence the smaller is the need for
the $r^2$ term which translates in a larger $r_A$.

\section{Conclusions}

Initially motivated by its attractive features from the point of view of quantum gravity, the HL
proposal has soon become one of the most investigated theories of gravity. We have here complemented recent
works on its cosmological consequences by addressing its impact on the gravitational potential. Contrary to
the claim in TC09, we have demonstrated that static spherically symmetric solutions other than the Schwartzschild\,-\,de 
Sitter one exist. As a consequence, we have found a modified gravitational potential made out of the Newtonian $1/r$ one
corrected by the addition of three further terms scaling as $r^2$, which corresponds to the effect of a cosmological constant,
and a quickly decreasing term, proportional to $1/r^4$. The importance of these two terms is parametrized in terms of two 
conveniently defined scaling radii, namely $(r_A, r_D)$, related to the couplings entering the HL Lagrangian. In order to 
consider astrophysically interesting situations, we have then developed a general formalism to compute the circular velocity curve 
provided the mass density and the mass function of the system are given. As an application, we have then evaluated the Milky Way rotation 
curve using as only source of the gravitational field a spheroidal truncated power\,-\,law bulge and a double exponential disc. It turns 
out that the modified rotation curve is unable to fit the data thus demonstrating that the HL theory can not play the role of an alterative 
solution to the missing mass problem.

It is worth noting that Mukohyama (2009) has shown that classical solutions to the infrared (IR) limit of the HL theory can mimic 
General Relativity plus cold dark matter so that one can argue that it should be possible for the HL theory to provide an effective 
dark matter halo in apparent contradiction with our finding. Actually, this is not the case. First, Mukohyama only refers to the HL 
IR limit so that the Lagrangian he has considered does not include any potential term, while we here explicitly include this term 
through the coupling parameters $g_i$. Second, the cold dark matter term comes out as a consequence of the global Hamiltonian constraint 
being less restrictive than the local one typically imposed in General Relativity. In our static spherically symmetric metric, the global 
Hamiltonian constraint reduces to Eq.(\ref{eq: ss2}). In order to find our solution, we then impose that the integrand in Eq.(\ref{eq: ss1}) 
identically vanishes so that we are actually converting the global constraint in a local one thus going back to a situation similar to the 
General Relativity case. As a consequence, we argue that the possibility to get a matter term as an integration constant is lost in our approach.

The next step in our analysis of the HL theory on galactic scale should naturally be the inclusion of a dark halo.
There are however some subtle issues making such a logical step forward not so easy to address. First, needless to
say, there are few hints on which the dark halo mass density profile should be. All the models used in literature are
motivated by the outcome of numerical simulations, such as, e.g., the popular NFW \cite{NFW97} and Einasto \cite{E65,Polls}, or evidences from rotation curve fitting, such as the isothermal sphere \cite{BT87} and the Burkert \cite{B95,BS00,BS01} profile. All the previous works implicitly assume the validity of the Newtonian potential so that they are no more valid in a modified framework as the one we are using here. As a consequence, we have therefore to explore a wide class of density profile able to mimic most of the models in literature to finally select the most empirically motivated one. Moreover, as Eq.(\ref{eq: rotcurvegen}) shows, we also need to know the dark matter mass function which is completely unknown. Should the dark matter be composed of pointlike particles all having the same mass, we can adopt a Dirac $\delta$ leaving the mass of the particle as an unknown or setting it according to some particle physics model. As a final consequence, the fit to the rotation curve should determine the three HL scaling parameters $(r_A, r_D)$ and ${\cal{N}}_{DM}$ halo parameters so that severe degeneracies among these ${\cal{N}}_{DM}$ quantities may take place. In order to reduce them, a better way should be to consider external galaxies rotation curves assuming that the MF is the same as the MW one. In such a way, we could take advantage of the data probing the full radial range and not only the outer disc as done here with the MW sample. Moreover, such a test can also probe the universality of the scaling radii $(r_A, r_D)$ thus providing a further mandatory test of the HL model.

As a final remark, we want to stress that, should such an analysis be successful, one has still to address a different issue. Let us suppose that we have indeed well fitted the rotation curves of a large sample of spiral galaxies thus determining a halo model and the values of $(r_A, r_D)$. One could then use Eqs.(\ref{eq: defhlpar}) and (\ref{eq: defradii}) to infer constraints on the  HL coupling parameters $(g_0, \ldots, g_8)$. Although we have thus five constraints (the two radii plus the two conditions imposed in the derivation of the potential) so being unable to determine all the eight quantities $(g_0, \ldots, g_8)$, one could nevertheless try to see whether the allowed region of the parameter space is consistent with what is inferred from cosmological analyses \cite{DS09} or Solar Sytem tests (see e.g. Iorio \& Ruggiero 2009 and references therein). It is worth noting that such an attempt should give a consistent picture of the HL framework on all physical scales thus allowing to draw a deeper insight into its viability. \\

{\it Acknowledgements.} We warmly thank Mauro Sereno for helpful comments on a preliminary version
of this manuscript. MC is supported by Regione Piemonte and Universit\`{a} degli Studi di Torino.

\end{document}